\documentclass{article}
\usepackage{amsmath}
\usepackage{amssymb}
\renewcommand{\ll}{\label} \newcommand{\begeq}{\begin{equation}}
\newcommand{\bea}{\begin{eqnarray}}
\newcommand{\eea}{\end{eqnarray}} \newcommand{\nn}{\nonumber}

\newcommand{\ca}{$C^*$-algebra} 
 \newcommand{\rep}{representation}

\newcommand{\Hs}{Hilbert space} 
 
\newcommand{\sta}{$\mbox{}^*$-algebra}

 \newcommand{\ovl}{\overline}
\newcommand{\wt}{\widetilde} \newcommand{\til}{\tilde}
\newcommand{\raw}{\rightarrow} \newcommand{\rat}{\mapsto}
\newcommand{\law}{\leftarrow}

\newcommand{\ot}{\otimes} 
\newcommand{\la}{\langle} \newcommand{\ra}{\rangle}
 \newcommand{\wed}{\wedge}

\newcommand{\x}{\times} 
 
\newcommand{\cin}{C^{\infty}} \newcommand{\cci}{C^{\infty}_c}

\newcommand{\inv}{^{-1}}

\newcommand{\al}{\alpha} \newcommand{\bt}{\beta}
\newcommand{\gm}{\gamma} \newcommand{\Gm}{\Gamma}

\newcommand{\zt}{\zeta} \newcommand{\et}{\eta}
 
\newcommand{\io}{\iota} 
\newcommand{\lm}{\lambda} \newcommand{\Lm}{\Lambda}
\newcommand{\rh}{\rho} \newcommand{\sg}{\sigma}
 \newcommand{\ta}{\tau} 
 \newcommand{\phv}{\varphi}
 \newcommand{\ps}{\psi} 
\newcommand{\om}{\omega} 
\newcommand{\A}{\mathfrak{A}} \newcommand{\B}{\mathfrak{B}}

 \newcommand{\g}{\mathfrak{g}}

 \newcommand{\CE}{{\mathcal E}}

   \newcommand{\CL}{{\mathcal L}}   
 \newcommand{\CM}{{\mathcal M}}

\newcommand{\C}{{\mathbb C}}

  \makeatletter
\newskip\tempskip \def\endproof{{\parfillskip24\p@ plus\@ne
fil\@@par}\tempskip\prevdepth
\ifdim\lastskip=\z@\tempskip\z@\else\vskip-\lastskip
\ifdim\tempskip>4\p@ \tempskip.5\tempskip \else \tempskip\z@\fi\fi
\nobreak\vskip-\baselineskip\vskip-\tempskip\noindent\hbox
to\hsize{\hfill
$\blacksquare$}\par\vskip\tempskip\vskip\abovedisplayskip\@doendpe}
\makeatother \makeatletter
\newskip\tempskip \def\endiproof{{\parfillskip24\p@ plus\@ne
fil\@@par}\tempskip\prevdepth
\ifdim\lastskip=\z@\tempskip\z@\else\vskip-\lastskip
\ifdim\tempskip>4\p@ \tempskip.5\tempskip \else \tempskip\z@\fi\fi
\nobreak\vskip-\baselineskip\vskip-\tempskip\noindent\hbox
to\hsize{\hfill
$\Box$}\par\vskip\tempskip\vskip\abovedisplayskip\@doendpe}
\makeatother \newcommand{\enp}{\endproof}

\newtheorem{theorem}{Theorem}[section]

\newtheorem{definition}[theorem]{Definition}

\newtheorem{proposition}[theorem]{Proposition}
\newtheorem{corollary}[theorem]{Corollary}
\newtheorem{remark}[theorem]{Remark}

\newcommand{\had}{|\Lm|^{1/2}}
\newcommand{\Me}{Morita equivalent}

\newcommand{\sym}{\mbox{\rm sym}}

\begin{document}
\pagestyle{plain}
\title{The Muhly--Renault--Williams theorem for Lie groupoids and
its classical counterpart}
\author{N.P. Landsman\thanks{Korteweg--de Vries Institute for Mathematics,
University of Amsterdam,
Plantage Muidergracht 24,
NL-1018 TV AMSTERDAM, THE NETHERLANDS, 
email \texttt{npl@science.uva.nl}}
\thanks{Supported by a Fellowship from the Royal Netherlands Academy
of Arts and Sciences (KNAW)}}
\date{\today}
\maketitle
\textbf{Abstract} A theorem of 
Muhly--Renault--Williams states that if two locally compact groupoids
with Haar system are Morita equivalent, then their associated
convolution \ca s are strongly \Me.  We give a new proof of this
theorem for Lie groupoids. Subsequently, we prove a counterpart of this
theorem in Poisson geometry: If two Morita equivalent Lie groupoids
are s-connected and s-simply connected, then their associated Poisson
manifolds (viz.\ the dual bundles to their Lie algebroids) are Morita
equivalent in the sense of P. Xu.
\section{Introduction}
There are two interesting constructions relating groupoids to \ca s.
Firstly, a locally compact groupoid $G$ with Haar system $\lm$ defines
an associated convolution \ca\ $C^*(G,\lm)$ \cite{Ren}.
Secondly, a Lie groupoid $G$ is intrinsically associated with
a convolution \ca\ $C^*(G)$ \cite{Con}. 

For example, for a Lie group
$G$ the \ca\ $C^*(G)$ is isomorphic to the usual convolution algebra
of $G$. For a manifold $G_1=G_0=M$ one has $C^*(M)\simeq C_0(M)$, and 
for a pair groupoid over a manifold $M$ one obtains 
the \ca\ of compact operators on $L^2(M)$.

Involving operator algebras, the above constructions could be said to
be of a ``quantum'' nature.  From that perspective, the Lie case has a
``classical'' counterpart, involving Poisson manifolds.
Namely, a Lie groupoid $G$ canonically defines a Poisson manifold
$A^*(G)$ \cite{Cou,CDW}, which is the dual vector bundle to the Lie
algebroid $A(G)$ associated with $G$ \cite{Pra,Mac}. Our interpretation
of the passage $G\mapsto A^*(G)$ as the classical analogue of $G\mapsto C^*(G)$
has been justified by an analysis showing that $C^*(G)$ is a deformation
quantization (in the sense of Rieffel) of the Poisson manifold $A^*(G)$
\cite{NPL3,NPLCMP,LR}.

For all four cases of locally compact groupoids, Lie groupoids, \ca s,
and Poisson manifolds there exists a notion of Morita equivalence; see
\cite{MRW}, \cite{Xu1}, \cite{Rie2}, and
\cite{Xu2}, respectively. 
A remarkable theorem of Muhly--Renault--Williams  \cite{MRW}
states that if two locally compact groupoids with Haar
system are Morita equivalent, then so are their associated convolution \ca s.

The fact that any Lie groupoid possesses a Haar system
\cite{Ram,NPL3} establishes 
  the corresponding result for Lie groupoids.  Nonetheless, we give a
  new proof of the Muhly--Renault--Williams theorem for Lie groupoids,
  which provides considerable insight into the situation. Our proof is
  not quite independent of the one in \cite{MRW}, for in the technical
  step of taking completions of various pre-Banach spaces we rely on
  certain ``hard''  results in the locally compact case \cite{MRW,Ren,Ren87}.

Subsequently, we prove a counterpart of this theorem in Poisson
geometry: If two Morita equivalent Lie groupoids are s-connected and
s-simply connected, then their associated Poisson manifolds (viz.\ the
dual bundles to their Lie algebroids) are Morita equivalent.  The
essential technical difficulty in the proof of this theorem, namely
the completeness of certain Poisson maps, is overcome by constructing
the pullback of the action of a Lie groupoid $G$ on a manifold $M$;
this is an action of the symplectic groupoid $T^*G$ on the cotangent
bundle $T^*M$. This construction also clarifies the definition of
$T^*G$ itself.  

\textbf{Acknowledgements} The author is indebted to the participants of
the Seminar on Groupoids 1999--2000, notably M. Crainic, K. Mackenzie,
I. Moerdijk, J. Mr\v{c}un, H. Posthuma, and J. Renault, for
discussions.

\section{The Muhly--Renault--Williams theorem for Lie groupoids}
\subsection{Statement of definitions and theorem}
\label{MRWLG}
Our generic notation for groupoids is that
$G_0$ is the base space of a  groupoid $G$, with source and target
maps $s,t:G_1\raw G_0$, 
multiplication $m:G_2\raw G_1$
(where $G_2= G_1 *^{s,t}_{G_0}G_1$),
inversion $I:G_1\raw G_1$, and object inclusion
$\io:G_0\hookrightarrow G_1$ (this inclusion map will often be  
taken for granted, in that $G_0$ is seen as a subspace of $G_1$). 

A Lie groupoid is a groupoid for which $G_1$ and $G_0$ are manifolds,
$s$ and $t$ are surjective submersions, and $m$ and $I$ are smooth.
It follows that $\io$ is an immersion, that $I$ is a diffeomorphism, 
that $G_2$
is a closed submanifold of $G_1\x G_1$, and that
for each $q\in G_0$ the fibers $s\inv(q)$ and $t\inv(q)$ are
submanifolds of $G_1$.
References on Lie groupoids that are relevant to the themes
in this paper include \cite{Mac,CDW,MiWe,CW,NPL3}.

Since they play a central role in  Morita theory for Lie groupoids,
we now define actions and bimodules of Lie groupoids (these notions
occur in a large number of papers, and probably go back to Ehresmann
and Haefliger, respectively). 
\begin{definition}\label{Gaction} 
\begin{enumerate}
\item
Let $G$ be a Lie groupoid and let $M\stackrel{\ta}{\raw} G_0$ be smooth.
  A left $G$-action on $M$ (more precisely, on $\ta$) is a smooth
map $(x,m)\mapsto xm$ from $G *^{s,\ta}_{G_0}M$ to $M$
(i.e., one has $s(x)=\ta(m)$), such
that $\ta(xm)=t(x)$, $xm=m$ for all $x\in G_0$, and $x(ym)=(xy)m$
whenever $s(y)=\ta(m)$ and $t(y)=s(x)$.  
\item
A right action of a Lie groupoid $H$ on $M\stackrel{\sg}{\raw} H_0$ is
a smooth map $(m,h)\mapsto mh$ from $M*^{t,\ta}_{H_0} H$ to $M$ that
satisfies $\sg(mh)=s(h)$, $mh=m$ for all $h\in H_0$, and 
$(mh)k=m(hk)$ whenever $\sg(m)=t(h)$ and $t(k)=s(h)$.
\item
A $G$-$H$ bibundle $M$  carries a left
$G$ action as well as a right $H$-action that commute. That is,
 one has
$\ta(mh)=\ta(m)$, $\sg(xm)=\sg(m)$, and $(xm)h=x(mh)$ for all $(m,h)\in
M*H$ and $(x,m)\in G*M$. On occasion, we simply write $G\raw M\law H$.

The maps $\ta$ and $\sg$ will sometimes be called the base maps of the
given actions.
\item
A left action of a Lie groupoid $G$ on $M\stackrel{\ta}{\raw} G_0$ is
called principal when $\ta$ is a surjective submersion, and the action
is free (in that $xm=m$ iff $x\in G_0$) and proper (that is, the map
$(x,m)\mapsto (xm,m)$ from $G*_{G_0}M$ to $M\x M$ is proper).

A similar definition applies to right actions.
\end{enumerate}
\end{definition}

We now recall the definition of Morita equivalence of groupoids used in
\cite{MRW}, adapted to the smooth (Lie) case \cite{Xu1}.
\begin{definition}\label{MEGR}
A $G$-$H$ bibundle $M$ between Lie groupoids is called
an equivalence bibundle when:
\begin{enumerate}
\item $M$ is  left and right principal;
\item One has $M/H\simeq G_0$ via $\ta$ and $G\backslash M\simeq H_0$
via $\sg$.
\end{enumerate}
Two Lie groupoids related by an equivalence bibundle are called Morita
equivalent.
\end{definition}

This concept of \Me\ will be related to that for \ca s \cite{Rie2}.
Since various equivalent definitions are possible \cite{RW},
we recall the one that will be used. For the notion of a Hilbert
$C^*$ module that occurs, see \cite{RW,NPL3}.
\begin{definition}\label{smeca}
\begin{enumerate}
\item
An $\A$-$\B$ Hilbert bimodule, where $\A$ and $\B$ are \ca s, is 
 a Hilbert $C^*$ module $\CE$ over $\B$, along with a nondegenerate
$\mbox{}^*$-homomorphism of $\A$ into the \ca\ 
of adjointable operators $\CL_{\B}(\CE)$.
\item
An equivalence Hilbert bimodule between
two \ca s $\A$ and $\B$ is 
an  $\A$-$\B$ Hilbert bimodule $\CM$ that in addition is a left
Hilbert $C^*$ module over $\A$, such that 
\begin{enumerate}
\item
The range of $\la\, ,\,\ra_{\B}$ is dense in $\B$;
\item 
The range of $\mbox{}_{\A}\la\, ,\,\ra$ is dense in $\A$;
\item
The $\A$-valued inner product is related to the $\B$-valued one by
\begin{equation}
\mbox{}_{\A}\la\ps,\phv\ra\zeta=\ps\la\phv,\zt\ra_{\B}, \label{rii}
\end{equation}
for all $\ps,\phv,\zeta\in\CM$.
\end{enumerate}
\item Two \ca s are called (strongly) \Me\ when there exists an
equivalence Hilbert bimodule between them.
\end{enumerate}
\end{definition}

The Muhly--Renault--Williams theorem for Lie groupoids then reads
\begin{theorem}\label{mrw}
If $G$ and  $H$ are \Me\ as Lie groupoids, then their 
associated \ca s $C^*(G)$ and
$C^*(H)$ are \Me\ as \ca s.
 \end{theorem}

As stated in the Introduction, this theorem follows from the corresponding
result for locally compact groupoids with Haar system \cite{MRW}.
The proof in \cite{MRW} consists of two steps. 

In the first step one sets up a pre-equivalence Hilbert bimodule
between $C^*(G,\lm)$ and $C^*(H,\mu)$, given a $G$-$H$ equivalence
bibundle $M$. Here a pre-equivalence Hilbert bimodule for \ca s $\A$
and $\B$ is defined as in Definition \ref{smeca}, with the difference
that $\A$ and $\B$ are replaced by dense subalgebras $\A_0$ and
$\B_0$, respectively, and the Hilbert $C^*$-module $\CE_0$ over $\B_0$
is not required to be complete. In the case at hand, one has
$\A_0=C_c(G,\lm)$, $B_0=C_c(H,\mu)$, and $\CE_0=C_c(M)$.

For the second step, see section \ref{2step} below.
In the Lie case, we have been able to replace the first step
of the proof of the locally compact case in \cite{MRW} by purely
differential geometric arguments. This requires some preparation.
\subsection{Half-densities on Lie groupoids}
\addcontentsline{toc}{subsection}{Half-densities on Lie groupoids}
Following \cite{Con}, we use the well-known formalism of
half-densities, for which we need to establish some notation.  Let $E$
be a vector bundle over a manifold $M$ with $n$-dimensional typical
fiber $E_m$. The bundle ${\sf A}(E)$ is defined as $\wed^n E$ minus
the zero section. This is a principal $\C^*$-bundle over $M$, whose
fiber at $m$ is the $n$-fold antisymmetric tensor product of $E_x$,
with $0$ omitted (here $\C^*$ is $\C\backslash \{0\}$, seen as a
multiplicative group).  For $\al\neq 0$, the bundle of $\al$-densities
$|\Lm|^{\al}(E)$ is the line bundle over $M$ associated to ${\sf
A}(E)$ by the \rep\ $z\rat |z|^{-\al}$ of $\C^*$ on $\C$.  Hence
sections of $|\Lm|^{\al}(E)$ may be seen as maps $\phv:{\sf
A}(E)\raw\C$ satisfying $\phv(zv)=|z|^{\al}\phv(v)$.  One has natural
(and obvious) isomorphisms
\bea
|\Lm|^{\al}(E)\ot |\Lm|^{\bt}(E)
& \simeq & |\Lm|^{\al+\bt}(E); \label{Lm1}\\
|\Lm|^{\al}(E\oplus F)& \simeq & |\Lm|^{\al}(E)\ot |\Lm|^{\al}(F).
\label{Lm2}
\eea

The point of this formalism is already evident in the simplest case,
where $E=TM$ and $\al=1$; for one may integrate sections of
$\cci(M,|\Lm|^1(TM))$ over $M$ without choosing a measure (even when
$M$ is non-orientable).  Similarly, using (\ref{Lm1}), $\int_M fg$
makes sense for $f,g\in \cci(M,|\Lm|^{1/2}(TM))$.
Generalizing this case, let $M\stackrel{\ta}{\raw}X$ be a fibration
for which $\ta$ is a surjective submersion, and let $T^{\ta}M$ be the
subbundle of $TM$ whose fibers are tangent to the fibers of $\ta$.
One may then integrate $f\in \cci(M,|\Lm|^1(T^{\ta}M))$, or
$fg$, where $f,g\in \cci(M,|\Lm|^{1/2}(T^{\ta}M))$, over
any fiber of $\ta$.
\subsection{The  category of principal $G$ bundles}
\addcontentsline{toc}{subsection}{The  category of principal $G$ bundles}
Recall the definition of a principal $G$ action (Definition
\ref{Gaction}). The collection of all such actions (or bundles) can be
made into a category, with unexpected choice of arrows.
This category greatly clarifies both the definition of a Lie groupoid 
\ca\ $C^*(G)$ and the proof of Theorem \ref{mrw}.
The construction of this category may be found in \cite{Ram}, which
contains further details.

Let $G$ be a Lie groupoid, and let $M\stackrel{\ta}{\raw}G_0$ 
be a principal left $G$-space. The $G$-action pulls back to a
$G$-action on $\had(T^{\ta}M)$, which thereby becomes a 
principal left $G$-space as well, and one has the isomorphism
\begeq
\cin_{c/G}(M,\had(T^{\ta}M))^G\simeq \cci(G\backslash 
M,G\backslash \had(T^{\ta}M)).\label{ramiso}
\end{equation}
Here the left-hand side consists of $G$-equivariant sections
(that is, $\phv(xm)=x\phv(m)$) with compact support up to $G$-translations.
As to the right-hand side, note that if  $E$ is a vector bundle over $X$
such that $E$ and $X$ are  principal left $G$-manifolds compatible 
with the bundle projection, then $G\backslash E$ is naturally a vector bundle
over $G\backslash X$.

In addition, let $N\stackrel{\sg}{\raw}G_0$ be a principal left
$G$-space. Then the fiber product $M*_{G_0}N$  is a principal left
$G$-space under the obvious action $x:(m,n)\mapsto (xm,xn)$. 
We now define the complex vector space
\begeq
(M,N)_G=\cin_{c/G}(M*_{G_0}N,\had(T^{\ta}M)\ot \had(T^{\sg}N))^G.
\end{equation}
In view of (\ref{Lm2}) and the obvious fact
\begeq
T^{\ta=\sg}_{(m,n)}(M*_{G_0}N)=T^{\ta}_m M\oplus T_n^{\sg}N \label{tasg}
\end{equation}
 for $(m,n)\in M*_{G_0}N$, one has the natural isomorphism
\begeq
(M,N)_G\simeq\cin_{c/G}(M*_{G_0}N,\had(T^{\ta=\sg}(M*_{G_0}N)))^G,
\end{equation}
which may clarify the meaning of $(M,N)_G$.

 The point is now that, given a third principal
left $G$-space $Q\stackrel{\rh}{\raw}G_0$, one has a pairing $(M,N)_G\x
(N,Q)_G\raw (M,Q)_G$, given by
\begeq
f*g(m,q)=\int_{\sg\inv(\ta(m))} f(m,\cdot)\otimes g(\cdot,q).
\end{equation}
This is well defined in view of (\ref{Lm1}) and subsequent paragraph;
note that $\ta(m)=\rh(q)$ by definition of $M*Q$.  Furthermore, one
has a map $*:(M,N)_G\raw (N,M)_G$, given by $f^*(n,m)=\ovl{\mbox{\rm
flip}[f(m,n)]}$, where flip$:V\ot W\raw W\ot V$ is given by flip$(v\ot
w)=w\ot v$. This map is involutive, in being antilinear and satisfying
$(f*g)^*=g^* *f^*$. It follows that the principal left $G$-manifolds
are the objects of a *-category whose arrows are the spaces $(M,N)_G$.
\subsection{The \ca\ of a Lie groupoid}
\addcontentsline{toc}{subsection}{The \ca\ of a Lie groupoid}
To define $C^*(G)$, note that $G\stackrel{t}{\raw} G_0$ is itself a
principal left $G$-manifold. Hence the vector space $(G,G)_G$ becomes a
\sta\ under the above multiplication $(G,G)_G\x (G,G)_G\raw (G,G)_G$ and
involution $(G,G)_G\raw (G,G)_G$. Equipped with a suitable norm,
$(G,G)_G$ is a pre-\ca\ whose completion is the groupoid \ca\
$C^*(G)$.  One has the natural isomorphisms (cf.\ (\ref{ramiso}) and
(\ref{klaasiso}) below)
\begin{eqnarray} 
(G,G)_G & \simeq & 
\cin_{c/G}(G*_{G_0}^{t,t}G,\had(T^{t=t}G*_{G_0}^{t,t}G)^G
\nn \\  & \simeq &
\cci(G/(G*_{G_0}^{t,t}G),G/\had(T^{t=t}(G*_{G_0}^{t,t}G)))
  \nn \\  & \simeq & \cci(G,\had(T^sG)\ot\had(T^tG)), 
\label{coniso}
\end{eqnarray}
so that $(G,G)_G$ is isomorphic with the convolution \sta\ defined by Connes
\cite{Con}.  The Lie groupoid \ca\ $C^*(G)$ is then the completion 
of $(G,G)_G$ in the norm
$\| f\|=\sup\{ \|\pi(f)\|\}$, where the supremum is taken over all
\rep s (on \Hs s) of $(G,G)_G$ (as a \sta)
 that are continuous with respect to the inductive limit topology on
 $(G,G)_G$. The existence of the supremum follows from results in the
 locally compact case, namely Prop.\ 4.2 in \cite{Ren87} and Prop.\
 II.1.7 in \cite{Ren}.  Here, as in the second step of the proof of
 the theorem at hand, it seems that taking completions necessarily
 involves the theory of locally compact groupoids with Haar system.

The second isomorphism in (\ref{coniso}) follows from the following,
more general case. For a principal left $G$-manifold $M$, one has the
diffeomorphism $$G/(G*_{G_0}^{t,\ta}M)\simeq M$$ under the map
\begeq
[x,m]_G\mapsto x\inv m; \label{xmm}
\end{equation}
 this is well defined since $t(x)=\ta(m)$ by definition of
 $G*_{G_0}^{t,\ta}M$, so that $(x\inv,m)\in G *_{G_0}^{s,\ta}M$.  As
 we have seen in (\ref{tasg}), one has $T_{(x,m)}G*_{G_0}^{t,\ta}M=
 T^t_xG\oplus T_m^{\ta}M$; the derivative of (\ref{xmm}) maps $T^t_xG$
 into $T^G_{x\inv m}M$ and maps $T_m^{\ta}M$ into $T_{x\inv
 m}^{\ta}M$. Here the vertical tangent space $T^G_mM$ consists of all
 vectors that are tangent to $G$ orbits. With (\ref{ramiso}) and
 (\ref{Lm1}) this yields the isomorphism
\bea
(G,M)_G& \simeq & \cin_{c/G}(G*_{G_0}^{t,\ta}M, \had(T^{t=\ta}G*_{G_0}^{t,\ta}M))^G
\nn \\ & \simeq &
\cci(M,\had(T^GM)\ot\had(T^{\ta}M)). \label{klaasiso}
\eea
The isomorphism (\ref{coniso}) is evidently a special case of this.
\subsection{Construction of the pre-equivalence Hilbert bimodule}
Analogous considerations for right actions lead to a right version
of (\ref{klaasiso}), viz.\ 
\bea
(M,H)_H & \simeq & \cin_{c/H}(M*_{H_0}^{\sg,s} H,
\had(T^{\sg=s}M*_{H_0}^{\sg,s} H))^H
\nn \\ & \simeq & 
\cci(M,\had(T^{\sg}M)\ot \had(T^HM)). \label{klaasiso2}
\eea

Condition 2 in Definition \ref{MEGR} implies
\begin{eqnarray}
T^{\sg}M & = & T^GM; \nn \\ 
 T^{\ta}M & = & T^HM.\label{MHG}
\end{eqnarray} 
By pullback, we obtain the isomorphism
\begeq
(G,M)_G\simeq (M,H)_H.
\end{equation} 

This gives us a pre-equivalence Hilbert bimodule $\CM_0$ between $(G,G)_G$ and
$(H,H)_H$, as follows:
\begin{itemize}
\item
Identifying $\CM_0$ with
$(M,H)_H$,  one obtains a right $(H,H)_H$
representation on $\CM_0=(M,H)_H$ from the pairing $(M,H)_H\x
(H,H)_H\raw (M,H)_H$; that is, for $\ps\in\CM_0$ and $B\in (H,H)_H$ one
puts $\ps B=\ps*B$. 
\item
Similarly, the map $\la
\ps,\phv\ra_{(H,H)_H}=\ps^**\phv$ maps from $(M,H)_H^*\x (M,H)_H= (H,M)_H\x
(M,H)_H$ into $(H,H)_H\subset C^*(H)$, providing an $(H,H)_H$-valued
inner product on $\CM_0$.
\item
 On the other hand, identifying $\CM_0$ with
$(G,M)_G$, one obtains a representation of $(G,G)_G$ on $\CM_0$ from
the pairing $(G,G)_G\x (G,M)_G\raw (G,M)_G$; for $\ps\in\CM_0$ and $A\in
(G,G)_G$ one puts $A\ps=A*\ps$.
\item
On the same identification, $\mbox{}_{(G,G)_G}\la
\ps,\phv\ra=\ps*\phv^*$ maps from $(G,M)_G\x (G,M)_G^*=(G,M)_G\x
(M,G)_G\raw (G,G)_G$, defining a $(G,G)_G$ valued inner product on
$\CM_0$.
\end{itemize}

The required algebraic properties, including (\ref{rii}), are trivial
consequences of the associativity of the $*$-product, and of the involutivity
of $\mbox{}^*$. Positivity of the inner products and density of their
images is also easily established using the method of P. Green
\cite{Gre} (section 2),
as in the locally compact case. Indeed, the Lie analogue
of Prop.\ 2.10 in \cite{MRW} may be directly proved for Lie groupoids
in the same way as for locally compact groupoids. See Lemmas 4.18-4.20
in \cite{Sta}.
\subsection{Taking completions}\label{2step}
One now has to show that our pre-equivalence Hilbert bimodule
can be completed. As is well known \cite{Rie1,RW}, a sufficient condition
for this to be possible is that for all $\ps\in\CM_0$ one has
the bounds $\la A\ps ,A\ps\ra_{\B_0}\leq
\| A\|^2\la \ps ,\ps\ra_{\B_0}$ for all $A\in \A_0$ and
$\mbox{}_{\A}\la\ps B,\ps B\ra\leq \| B\|^2\mbox{}_{\A}\la\ps,\ps \ra$
for all $B\in\B_0$. That these bounds are satisfied in our groupoid
situation follows from two deep results of Renault, 
viz.\ Prop.\ 4.2 in \cite{Ren87} and Prop.\ II.1.7 in \cite{Ren}.

Thus we have been unable to modify the final stage of the proof of
\cite{MRW} by specific Lie groupoid arguments, but given the fact that
taking completions necessarily abandons the smooth setting, it seems
doubtful that such arguments exist.
\section{A classical analogue of the Muhly--Renault--Williams theorem
for Lie groupoids}
\label{MRWCL}
\subsection{Statement of definitions and theorem}
We recall the passage from a Lie group to its
Lie algebra \cite{Pra,Mac}
\begin{remark}\label{GAG}
A Lie groupoid $G$ defines a Lie algebroid $A(G)$ over $G_0$, as follows.
\begin{enumerate}
\item
The vector bundle $A(G)$ over $G_0$ is the 
kernel of $Tt$ (the
derivative of the target projection $t:G\raw G_0$) restricted (or
pulled back) to $G_0$; hence 
\begeq
A(G)=\ker(Tt)_{|G_0}.\label{defAG}
\end{equation}
  Accordingly, the
bundle projection is given by $s$ or $t$ (which coincide on $G_0$).
\item
The anchor is given by $a=Ts$ (restricted to $A(G)$).
\item
Identifying a section of $A(G)$ with a left-invariant
vector field on $G_1$, the Lie bracket $[\, ,\,]_{A(G)}$ is
given by the commutator of vector fields on $G_1$.
\end{enumerate}\end{remark}

For example, $TQ$ is the Lie algebroid of the pair groupoid $Q\x Q$,
and the Lie algebra $\g$ of a Lie group is its Lie algebroid.

Note that, since $\ker(Tt)_{|G_0}$ is a complement to $T(\io(G_0))$,
the Lie algebroid $A(G)$ is isomorphic to the normal bundle $\til{A}(G)$ of
the embedding $\io:G_0\hookrightarrow G$. This isomorphism endows
$\til{A}(G)$ with the structure of a Lie algebroid as well, isomorphic
to $A(G)$, and this alternative version is often called the Lie algebroid
of $G$, too (cf., e.g., \cite{CDW}).

One part of the connection between Lie algebroids and Poisson manifolds
is laid out by the following result \cite{Cou,CDW}.
\begin{proposition}\label{LAPM}
The dual vector bundle $E^*$ to a Lie algebroid $E$ has a canonical Poisson
structure that is linear. Conversely, any vector bundle with
a linear Poisson structure is dual to a Lie algebroid.
This establishes a categorical equivalence between linear Poisson structures
on vector bundles and Lie algebroids.

In particular, the dual vector bundle $A^*(G)$ of the Lie algebroid $A(G)$ of
 a Lie groupoid $G$, as well as the dual bundle $\til{A}^*(G)$ of
$\til{A}(G)$ (which is isomorphic to $A^*(G)$)
accordingly become  Poisson manifolds.
\end{proposition}

Here linearity means that the Poisson bracket of two linear functions
is linear; a function on $E^*$ is, in turn, called
linear when it is linear on each fiber. Each section $\sg$ of $E$
defines such a function $\til{\sg}$ in the obvious way. Also, each
$f\in\cin(Q)$ (where $Q$ is the base of $E$) trivially defines
$\til{f}\in\cin(E^*)$. The Poisson bracket on $E^*$ is then determined
by the following special cases:
\begin{eqnarray} \{\til{f},\til{g}\} & = & 0; \ll{pblieoid1} \\ 
\{\til{\sg},\til{f}\} & =
&  \widetilde{(a_*\sg) f}; \ll{pblieoid2} \\ 
\{\til{\sg}_1,\til{\sg}_2\}
& = &  \wt{[\sg_1,\sg_2]_{E}}. \ll{pblieoid3} 
\end{eqnarray}

These formulae show quite clearly how the data of a Lie algebroid
determine the Poisson structure, which is of a special kind.
For example, for a Lie group $G$ the Poisson manifold $A^*(G)$ is
just the dual of the Lie algebra of $G$, equipped with the usual Lie--Poisson
structure. For a manifold $G_1=G_0=M$ one finds $A^*(G)=M$ with zero
Poisson bracket, and for a pair groupoid $G_1=M\x M$ one obtains
$A^*(G)=T^*M$ with the canonical (symplectic) Poisson structure.
\begin{remark}\label{AAtil}
Note that $\til{A}^*(G)$ is the subbundle of $T^*G$ consisting of
1-forms over $G_0$ that annihilate $TG_0\subset TG_{| G_0}$.
The isomorphism $\til{A}^*(G)\simeq A^*(G)$ arises as follows:
for each $q\in G_0$ one has a decomposition
\begeq
T_qG=A_qG\oplus T_q G_0; \label{TGdec}
\end{equation}
cf.\ (\ref{defAG}).
Hence $\al_q\in A^*_q(G)$ defines $\til{\al}_q\in\til{A}_q^*(G)\subset T^*_qG$
by putting $\til{\al}_q=\al_q$ on $A_q(G)$ and $\til{\al}_q=0$ on
$T_q G_0$. Conversely, $\til{\al}_q\in\til{A}_q^*(G)$ defines
$\al_q\in A^*_q(G)$ by restricting it to $A_q(G)\subset T_qG$.
\end{remark}

The theory of Morita equivalence of Poisson manifolds was initiated by
Xu \cite{Xu2}, who gave the following definition.
\begin{definition}\label{MEPM}
\begin{enumerate}
\item
A symplectic bimodule $Q\stackrel{q}{\law} S
 \stackrel{p}{\raw}P$ for two Poisson manifolds $P$, $Q$ consists of a
 symplectic space $S$ with complete Poisson maps $p:S\raw P^-$ and
 $q:S\raw Q$, such that $\{p^*f,q^*g\}=0$ for all $f\in\cin(P)$ and
 $g\in\cin(Q)$.
\item
A symplectic bimodule $Q\law S\raw P$ is called an equivalence
symplectic bimodule when:
\begin{enumerate}
\item
The maps $p:S\raw P$ and $q:S\raw Q$ are surjective submersions;
\item The level sets of $p$ and $q$ are connected and simply connected;
\item The foliations of $S$ defined by the
levels of $p$ and $q$ are mutually symplectically orthogonal (in that
the tangent bundles to these foliations are each other's
  symplectic orthogonal complement).
\end{enumerate}
\item Two Poisson manifolds are called  \Me\ when there exists an
equivalence symplectic bimodule between them.
\end{enumerate}
\end{definition}

Our ``classical'' analogue of Theorem \ref{mrw} is now as follows.
\begin{theorem}\label{mrwcl}
Let $G$ and $H$ be s-connected and s-simply connected Lie groupoids,
with associated Poisson manifolds $A^*(G)$ and $A^*(H)$ 
 If $G$ and $H$ are \Me\ as Lie groupoids, then
$A^*(G)$ and $A^*(H)$ are Morita equivalent as Poisson manifolds; cf.\ 
Definition \ref{MEPM}.
\end{theorem}

The outline of the proof is as follows. Given a $G$-$H$ bibundle $M$
implementing the Morita equivalence of $G$ and $H$ (see Definition
\ref{MEGR}), we equip $S=T^*M$ with the structure
of an $A^*(G)$-$A^*(H)$ symplectic bimodule that satisfies all conditions in
Definition \ref{MEPM}. This involves two constructions that are 
interesting in their own right, which are the subject of sections
\ref{mmga} and \ref{pbga}.
\subsection{The momentum map for Lie groupoid actions}\label{mmga}
\addcontentsline{toc}{subsection}{The momentum map for Lie groupoid actions}
The basic construction is valid in more generality than our situation needs.
\begin{proposition}\label{premrw}
A left action of a Lie groupoid $G$ on a manifold $M$ defines a
complete Poisson map $J_L:T^*M^-\raw A^*(G)$ (called the momentum map of the
$G$ action).  Here $A^*(G)$ and $T^*M=A^*(M\x M)$ carry the Poisson structure
defined in Proposition \ref{LAPM} (which induces the canonical one on
$T^*M$).

Similarly, A right action of a Lie groupoid $H$ on $M$ defines a
complete Poisson map $J_R:T^*M\raw A^*(H)$.
\end{proposition}

Except for the completeness of $J_L,J_R$,
the proof is a straightforward generalization of the case where $G$ and
$H$ are Lie groups.
 The $G$-action leads to a map $\xi^L:A(G)\raw TM$, $X\mapsto\xi^L_X$, 
for which $\ta_{M\raw G_0}\circ\ta_{TM\raw M}(\xi^L_X)=\ta_{A(G)\raw G_0}(X)$.
With 
\begeq
X=\frac{d\gm(\lm)}{d\lm}_{|\lm=0}\in\pi\inv(q), \label{Xq}
\end{equation}
 $q\in G_0$, where, by definition 
of the Lie algebroid $A(G)$, one has 
\begeq
t(\gm(\lm))=t(\gm(0))=q \label{tq}
\end{equation}
 for all $\lm$,
this map is given by \cite{NPL3}
\begeq
\xi^L_X(m)=-\frac{d}{d\lm} \gm(\lm)\inv m_{|\lm=0}. \label{vert}
\end{equation}
Here $\ta(m)=q$.
Note that $\xi^L_X\in T_mM$, since $\gm(0)\in G_0$ by definition
of the Lie algebroid, and $\gm(0)m=m$ by definition of a groupoid action.
This yields our momentum map  by
\begeq
\langle J_L(\theta),X\rangle=\langle\theta,\xi^L_X\rangle.\label{defJG}
\end{equation}
One then checks that $J_L:T^*M\raw A^*(G)$ is an anti-Poisson map, so
that $J_L:T^*M^-\raw A^*(G)$ is a Poisson map, as follows.
As before, we write $\ta=\ta_{M\raw G_0}$.

For $f\in\cin(G_0)$ one has $J_L^*\til{f}=\hat{f}$,
where $\hat{f}=f\circ\ta\circ\ta_{T^*M\raw M}$, so that
$$\{J_L^*\til{f},J_L^*\til{g}\}_{T^*M}=\{\hat{f},\hat{g}\}_{T^*M}=0
=J_L^*\{\til{f},\til{g}\}_{A^*(G)}
$$
by (\ref{pblieoid1}).

For a section $\sg$ of $A(G)$, which we take to be of the form
$\sg(q)=X(q)$, as in (\ref{Xq}), with $q$-dependent curves $\gm_q(\lm)$,
one obtains a vector field $\xi_{\sg}^L$, in terms of which
 $J_L^*\til{\sg}=\sym(\xi_{\sg}^L)$. Here $\sym(\xi)\in\cin(T^*M)$
denotes the symbol of a vector field $\xi$ on $M$. The canonical
Poisson bracket on $T^*M$ satisfies
\begeq
\{\sym(\xi),h\}_{T^*M}(\theta_m)=\xi h(m) \label{sym1}
\end{equation}
 for $h\in\cin(M)$, so that
\begin{eqnarray*}
\{J_L^*\til{\sg},J_L^*\til{f}\}_{T^*M}(\theta_m)& = & \xi_{\sg}^L\hat{f}(m)=-
\frac{d}{d\lm} f(\ta[\gm_q(\lm)\inv m])_{|0}=-
\frac{d}{d\lm} f(s(\gm_q(\lm)))_{|0}\\
& =& -(Ts)(X(q))f(q)=
-(a_*\sg)f(q)=-J_L^*\{\til{\sg},\til{f}\}_{A^*(G)}(\theta_m),
\end{eqnarray*}
where $q=\ta(m)$.  Here we used  (\ref{pblieoid2}) and Remark \ref{GAG}.2.

Finally, using Remark \ref{GAG}.3, the property
\begeq
\{\sym(\xi),\sym(\et)\}_{T^*M}=\sym([\xi,\et]),\label{sym2}
\end{equation}
and (\ref{pblieoid3}),
one  proves that 
$$
\{J_L^*\til{\sg_1},J_L^*\til{\sg_2}\}_{T^*M}=-
J_L^*\{\til{\sg_1},\til{\sg_2}\}_{A^*(G)}.
$$

Since the differentials of the functions in question span $T^*(A^*(G))$,
this proves that $J_L:T^*M^-\raw
A^*(G)$ is a Poisson map. 

For the right $H$-action we define $J_R:T^*M\raw A^*(H)$ by
\begeq
\left\langle J_R(\theta_m),\frac{dh(\lm)}{d\lm}_{|\lm=0}\right\rangle=
\left\langle\theta_m,\frac{d m h(\lm)}{d\lm}_{|\lm=0}
\right\rangle,\label{JRE}
\end{equation}
where $h(\lm)\in t_H\inv(\sg(m))$, so that its tangent vector at 0 lies
in $A_{\sg(m)}H$, and the expression $mh(\lm)$ is defined.
This may be shown to be a Poisson map by essentially the same computations
as for $J_L$. 

The completeness of $J_L$ and $J_R$ will be proved in 
 section \ref{pbga}.
\enp

The corresponding momentum
maps $\til{J}_L:T^*M^-\raw \til{A}^*(G)$ and $\til{J}_R:T^*M\raw \til{A}^*(H)$
(cf.\ Remark \ref{AAtil}) arise in the obvious way, by extending
the given expression by 0 on $TG_0$. However, it is instructive
to rewrite $\til{J}_L$. 
Instead of (\ref{TGdec}), we now use the decomposition
$TG_{|G_0}=\ker(Ts)_{|G_0}\oplus TG_0$. Relative to this,
a vector
$dx/d\lm_{|0}\in \ker(Tt)$, with $\gm(0)=q\in G_0$,  decomposes as
\begeq
\frac{d\gm(\lm)}{d\lm}_{|\lm=0}=-\frac{d\gm(\lm)\inv}{d\lm}_{|\lm=0}
+\frac{ds(\gm(\lm))}{d\lm}_{|\lm=0}.
\end{equation}

Hence on $\ker(Ts)_{|G_0}\subset TG_{|G_0}$ we simply have
\begeq
\left\langle \til{J}_L(\theta_m),\frac{dz(\lm)}{d\lm}_{|\lm=0}\right\rangle
=
\left\langle\theta_m,\frac{dz(\lm)m}{d\lm}_{|\lm=0}\right\rangle.
\end{equation}
Here $z(\lm)$ lies in the $s$-fiber above $\ta(m)\in G_0$, so that
the right-hand side is defined. Compare this with (\ref{defJG}), which
may be written as
\begeq
\left\langle J_L(\theta_m),\frac{d\gm(\lm)}{d\lm}_{|\lm=0}\right\rangle
= -\left\langle\theta_m,\frac{d\gm(\lm)\inv
m}{d\lm}_{|\lm=0}\right\rangle,\label{Jrev}
\end{equation}
where $\gm(\lm)$ lies in the $t$-fiber above $\ta(m)$.
\begin{corollary}\label{MRWcl}
Let $G$ and $H$ be Lie groupoids, and let $M$ be a $G$-$H$ bibundle.
 Then there exist maps
$J_L,J_R$ for which 
\begeq A^*(G)\stackrel{J_L}{\longleftarrow} T^*M^-
\stackrel{J_R}{\longrightarrow}A^*(H)
\end{equation}
is a symplectic bimodule.
\end{corollary}

The definition of a groupoid bibundle easily implies that the last
condition in Definition \ref{MEPM}.1 is met:
Firstly, $$
\{J^*_L\til{f},J^*_R\til{g}\}_{T^*M}=\{\hat{f},\check{g}\}_{T^*M}=0,
$$ where
$\check{g}=g\circ\sg\circ\ta_{T^*M\raw M}$. Secondly, using (\ref{sym1}),
one has
$$
\{J^*_L\til{\sg},J^*_R\til{g}\}_{T^*M}=\xi^L_{\sg}\check{g}=0,
$$
since $\sg:M\raw H_0$ is $G$-invariant. Similarly,
$$
\{J^*_L\til{f},J^*_R\til{\sg}\}_{T^*M}=-\xi^R_{\sg}\hat{f}=0,
$$
since $\ta:M\raw G_0$ is $H$-invariant.
Finally, using (\ref{sym2}) and the fact that the $G$ and $H$ actions
on $M$ commute, one computes
$$
\{J^*_L\til{\sg_1},J^*_R\til{\sg_2}\}_{T^*M}=\sym([\xi^L_{\sg_1},
\xi^R_{\sg_2}])=\sym(0)=0.
$$
Checking Poisson commutativity for the given functions suffices.
\enp
\subsection{The cotangent bundle of a Lie groupoid}
\addcontentsline{toc}{subsection}{The cotangent bundle of a Lie groupoid}
In order to prove completeness of the maps $J_L$ and $J_R$, we will need
the cotangent bundle of a Lie groupoid \cite{CDW}.
We here reinterpret their source and target maps  in terms of
the momentum maps $J_L$ and $J_R$ of the preceding section.
\begin{proposition}\label{TsGo}
The cotangent bundle $T^*G^-$ of a Lie groupoid $G$ becomes a
symplectic groupoid over $\til{A}^*(G)$ in the following way (we here
work with $\til{A}^*(G)$ rather than $A^*(G)$ in order to facilitate the
use of \cite{CDW}).  Consider $G$ as a $G$-$G$ bibundle in the obvious
way.  The source map $\til{s}:T^*G\raw A^*(G)$ is given by
$\til{s}=\til{J}_R$, the target is $\til{t}=\til{J}_L$, the object
inclusion map is $\til{A}^*(G)\hookrightarrow T^*G$, inversion is
$\til{I}=-I^*$, and multiplication is defined as follows.

First note that, by definition of $\til{J}_L$ and $\til{J}_R$, one has
$\til{s}(\al_x)\in \til{A}^*_{s(x)}(G)$ and $\til{t}(\bt_y)\in
\til{A}^*_{t(x)}(G)$.  Hence the condition $(\al_x,\bt_y)\in T^*G_2$
implies $(x,y)\in G_2$.  As in \cite{CDW}, one shows that the former
condition implies that there exists a (necessarily unique)
$\gm_{xy}\in T^*_{xy}G$ such that
\begeq
\al_x(X)+\bt_y(Y)=\gm_{xy}(T_{(x,y)}m(X,Y))\label{mTsG}
\end{equation} 
for all $(X,Y)\in T_{(x,y)}G_2$, and this $\gm_{xy}$ in fact lies in
$\til{A}_{xy}^*(G)$.  The multiplication $\til{+}$ in $T^*G$ is then
given by 
\begeq
\al_x\til{+}\bt_y=\gm_{xy}.\label{deftilplus}
\end{equation} 
\end{proposition}
\subsection{The pullback of a Lie groupoid action}\label{pbga}
\addcontentsline{toc}{subsection}{The pullback of a Lie groupoid action}
We will prove that $J_L$ and $J_R$ are complete by constructing 
symplectic actions of the symplectic groupoids $T^*G$ and $T^*H$
(cf.\ Proposition \ref{TsGo}) on $T^*M$ with base maps $J_L$ and $J_R$,
respectively, Completeness then follows from Thm.\ 3.1 in \cite{Xu2},
stating that the  base map of a symplectic groupoid action is
automatically complete.

The following theorem covers the general situation. It generalizes
Ex.\ 3.9 in \cite{MiWe} from groups to groupoids, and its corollary
of completeness generalizes Lemma 3.1 in \cite{Xu3}.
\begin{theorem}\label{coJ}
Let $G$ be a Lie groupoid acting on a manifold $M$, with
associated momentum map $J_L:T^*M^-\raw A^*(G)$ 
(cf.\ Proposition \ref{premrw}).

There exists a symplectic action of $T^*G^-$ 
(cf.\ Proposition  \ref{TsGo}) on $T^*M^-$ with base map $J_L$. In particular,
$J_L$ is complete. 
\end{theorem}


Take $\al_x\in T^*_xG$ and $\theta_m\in T^*_mM$ such that
$\til{s}(\al_x)=J_L(\theta_m)$. According to (\ref{Jrev}) 
and Proposition \ref{TsGo}, using  (\ref{JRE}) applied to the case $M=G$,
this condition implies $s(x)=\ta(m)$, and otherwise reads
\begeq
\left\langle\al_x,\frac{dx\gm(\lm)}{d\lm}_{|\lm=0}\right\rangle=-
\left\langle\theta_m, \frac{d\gm(\lm)\inv m}{d\lm}_{|\lm=0}\right\rangle.\label{isis}
\end{equation}

Here $\gm(\lm)\in t\inv(s(x))$. We now define $\al_x\cdot\theta_m\in
T^*_{xm}M$ as follows. Given $dn/d\lm_{|0}\in T_{xm}M$,
one picks a $t$-cover $g(\cdot)$ in $G$ of the curve $\ta(n(\cdot))$ in $G_0$;
that is, one has $g(0)=x$ and $t(g(\lm))=\ta(n(\lm))$.
We then put 
\begeq
\left\langle \al_x\cdot\theta_m, \frac{dn(\lm)}{d\lm}_{|\lm=0}\right\rangle=
\left\langle\theta_m, \frac{dg(\lm)\inv n(\lm)}{d\lm}_{|\lm=0}\right\rangle
+\left\langle\al_x,\frac{dg(\lm)}{d\lm}_{|\lm=0}\right\rangle.\label{defTGA}
\end{equation}

The arbitrariness in the choice of $g(\cdot)$ is immaterial because of
(\ref{isis}). To see this, one replaces $g(\lm)$ by a curve
$g(\lm)h(\lm)$ with the same properties, finding that $h$ drops out
of (\ref{defTGA}). Equivalently, we may write (\ref{defTGA}) as
\begeq
\left\langle\al_x\cdot\theta_m,\xi_{xm}\right\rangle=\left\langle\theta_m, 
T_{(x\inv,xm)}\phv(
T_xI(\et_x)+\xi_{xm})\right\rangle +
\left\langle\al_x,\et_x\right\rangle.\label{TGAbis}
\end{equation}
Here $\phv:G*_{G_0}^{s,\ta}M\raw M$ is the given $G$-action, and
$\et_x\in T_x$ covers $T\ta(\xi_{xm})$ under $t$, i.e., 
$T_xt(\et_x)=T_{xm}\ta(\xi_{xm})$.  The arbitrariness in $\et_x$
is a vector in $\ker(Tt)$, which drops out of (\ref{TGAbis})
because of (\ref{isis}) and the fact that $\ker(Tt)$ is spanned by
vectors of the form occurring on the left-hand side of that equation.

We now check that $J_L(\al_x\cdot\theta_m)=\til{t}(\al_x)$.
Evaluating both sides on a vector $d\gm/d\lm_{|0}$, this
 condition may be rewritten as
\begeq
\left\langle\al_x\cdot\theta_m,
\frac{d\gm(\lm)\inv xm}{d\lm}_{|\lm=0}\right\rangle=
\left\langle\al_x, 
\frac{d\gm(\lm)\inv x}{d\lm}_{|\lm=0}\right\rangle.\label{tcon}
\end{equation}

To compute the left-hand side, we take $n(\lm)=\gm(\lm)\inv xm$ and
$g(\lm)=\gm(\lm)\inv x$ in (\ref{defTGA}). The first term on the
right-hand side of (\ref{defTGA}) then vanishes, and the second term
equals the right-hand side of (\ref{tcon}).

Next, we verify that 
\begeq
\al_x\cdot (\bt_y\cdot\theta_m)=(\al_x\til{+}\bt_y)\cdot\theta_m, 
\label{abth}
\end{equation}
whenever defined. We compute the left-hand side from (\ref{defTGA})
as
\begin{eqnarray*}
\left\langle \al_x(\bt_y\cdot\theta_m), 
\frac{dn(\lm)}{d\lm}_{|\lm=0}\right\rangle & = &
\left\langle\bt_y\cdot\theta_m, 
\frac{dg(\lm)\inv n(\lm)}{d\lm}_{|\lm=0}\right\rangle
+\left\langle\al_x,\frac{dg(\lm)}{d\lm}_{|\lm=0}\right\rangle \\
& = &
\left\langle\theta_m, 
\frac{dh(\lm)\inv g(\lm)\inv n(\lm)}{d\lm}_{|\lm=0}\right\rangle \\
& + & \left\langle\bt_y,\frac{dh(\lm)}{d\lm}_{|\lm=0}\right\rangle
+\left\langle\al_x,\frac{dg(\lm)}{d\lm}_{|\lm=0}\right\rangle.
\end{eqnarray*}

Here $g$ is as specified after (\ref{isis}), and $h$ is such that
$h(0)=y$ and $t(h(\lm))=\ta(g(\lm)\inv n(\lm))=s(g(\lm))$.
The right-hand side of (\ref{abth}) is computed as follows: as the
$t$-cover $\tilde{g}(\cdot)$ of $n(\cdot)$ satisfying
$\til{g}(0)=xy$ and $t(\til{g}(\lm))=\ta(n(\lm))$ we may use
$\til{g}(\lm)=g(\lm)h(\lm)$. Eq.\ (\ref{abth}) is then immediate from
(\ref{mTsG}) and (\ref{deftilplus}).

Finally, we show that elements of $(T^*G)_0$ act trivially on $T^*M$.
According to the definition of the groupoid structure of $T^*G$, a
unit $\al_x\in T^*_xG$ satisfies $x\in G_0$ and $\al_x|T_xG_0=0$.
The former condition implies that in (\ref{defTGA}) we may take
$g(\lm)=\ta(n(\lm))$, so that $g(\cdot)\subset G_0$. The second condition
then implies that the second term on the right-hand side of
 (\ref{defTGA}) vanishes, whereas the first term is
$\langle\theta_m,dn/d\lm_{|0}\rangle$; this is because $g(\lm)\inv n(\lm)=
n(\lm)$, since $g(\lm)\inv\in G_0$.

It is routine to check that the $T^*G$ action on $T^*M$ is smooth.
That it is symplectic may be verified from a local computation
showing that the graph of $(\al_x,\theta_m)\mapsto \al_x\cdot\theta_m$
is coisotropic in $T^*G^-\x T^*M^-\x T^*M$. An easy dimensional count
then implies that it is Lagrangian.

For the final claim in Theorem \ref{coJ}, see the beginning of this
section.\enp

When $G$ is a Lie group, we may choose $\eta_x=0$ in (\ref{TGAbis})
to compute
$$
\left\langle\al_x\cdot\theta_m,\xi_{xm}\right\rangle=\left\langle\theta_m, 
T_{(x\inv,xm)}\phv(\xi_{xm})\right\rangle =
\left\langle \phv^*_{x\inv}\theta_m,\xi_{xm}\right\rangle,
$$
where $\phv_x:m\mapsto xm$ is the $G$ action on $M$.
Hence $\al_x\cdot\theta_m=\phv^*_{x\inv}\theta_m$, and our
$T^*G$ action on $T^*M$ is just the pullback of the $G$ action on $M$.
Also see \cite{MiWe}.
\subsection{Proof of Theorem \ref{mrwcl}}
\addcontentsline{toc}{subsection}{Proof of Theorem \ref{mrwcl}}
Let us now specialize Corollary \ref{MRWcl} to the situation of
Theorem \ref{mrwcl}, where the bibundle $M$ satisfies the conditions
in Definition \ref{MEGR}. Condition 1 in the latter easily implies
that the maps $p=J_R$ and $q=J_L$ satisfy condition 1 in Definition
\ref{MEPM}. To prove condition 2 (for $q$ to be concrete), we first note that
the set $J_L\inv(\al)$ by construction is a sub vector bundle of the
restriction of $T^*M$ to $M_{\al}=\ta\inv(\pi^{(*)}(\al))\subset M$
(where $\pi^{(*)}:A^*(G)\raw G_0$ is the bundle projection of $A^*(G)$
dual to $\pi:A(G)\raw G_0$; cf.\ Definition \ref{GAG}).  By property 2
in Definition \ref{MEGR}, the latter set is an $H$-orbit, and by
property 1 in Definition \ref{MEGR} (for $H$) and Definition
\ref{Gaction}.2 this orbit is diffeomorphic to $t_H\inv(\sg(m))$,
where $m\in M_{\al}$ (for a different choice $m'\in M_{\al}$ one has
$m'=mh$ for some $h\in t_H\inv(\sg(m))$, and then $t_H\inv(\sg(m'))$
is diffeomorphic with $t_H\inv(\sg(m))$ through $k\mapsto hk$). Hence,
by assumption in Theorem \ref{mrwcl}, condition 2 in Definition
\ref{MEPM} holds for $q$. An isomorphic argument with $G$ and $H$
interchanged proves this condition for $p$.

Finally, to prove condition 3 in Definition \ref{MEPM}, proceed as
follows. First compute the tangent spaces $TJ_L\inv$ to the fibers of
$J_L$: one has $X\in T_{\al}J_L\inv$ iff $X(J_L^* f)=0$ for all
$f\in\cin(A^*(G))$. Splitting $f$ into the types $\til{f}$ and
$\til{\sg}$ discussed earlier, and assuming $X=X_g$ is a 
Hamiltonian vector field (allowed, as $T^*M$ is symplectic),
this implies $g\in\cin(M)^G$ or $g=\sym(\xi)$, where
$\xi\in\Gm(T^{\ta}M)^G$. Similarly, $Y_h\in T_{\al}J_R\inv$ 
when $h\in\cin(M)^H$ or $h=\sym(\eta)$, where
$\eta\in\Gm(T^{\sg}M)^H$.
The inclusion $T_{\al}J_R\inv\subseteq (T_{\al}J_L\inv)^{\perp}$
now follows as in the proof of Corollary \ref{MRWcl} (using the basic
fact that $\om(X_f,X_g)=\{f,g\}$). 
The opposite inclusion follows from the crucial information (\ref{MHG}).
\enp

Let us finally note that the above proof has the following
reinterpretation.  By a remarkable theorem of Dazord \cite{Daz} and Xu
\cite{Xu2}, if $P$ is an integrable Poisson manifold with s-connected
and s-simply connected symplectic groupoid $\Gm(P)$, any complete
Poisson map $J:S\raw P$ defines a symplectic action of $\Gm(P)$ on
$S$, and vice versa. Another theorem of Xu \cite{Xu2} states that two
Poisson manifolds $P$ and $Q$ are Morita equivalent iff their
associated s-connected and s-simply connected symplectic groupoids
$\Gm(P)$ and $\Gm(Q)$ are Morita equivalent. Applied to the case at
hand, we have $P=A^*(G)$, $Q=A^*(H)$, $\Gm(P)=T^*G^-$, and
$\Gm(Q)=T^*H^-$.  Our proof shows that $T^*M^-$ is a symplectic
equivalence bimodule between $T^*G^-$ and $T^*H^-$, establishing their
Morita equivalence as symplectic groupoids. Hence their associated
Poisson manifolds $A^*(G)$ and $A^*(H)$ are Morita equivalent as well.

\end{document}